\documentclass[twocolumn,aps,10pt,prl,showpacs,showkeys,]{revtex4}
\usepackage[dvips]{graphicx}

\begin{document}
\title{NiO - from first principles}
\author{R. J. Radwanski}
\affiliation{Center of Solid State Physics, S$^{nt}$Filip 5,
31-150 Krakow, Poland\\
Institute of Physics, Pedagogical University, 30-084 Krakow,
Poland}
\author{Z. Ropka}
\affiliation{Center of Solid State Physics, S$^{nt}$Filip 5,
31-150 Krakow, Poland} \homepage{http://www.css-physics.edu.pl}
\email{sfradwan@cyf-kr.edu.pl}

\begin{abstract}
We have calculated from first-principles the octupolar
interactions of the Ni$^{2+}$ ion in NiO, which gives the
theoretical basis for the ionic description of properties of NiO
with fully localized strongly-correlated eight $d$ electrons. A
failure of the up-now first principles ionic calculations for NiO
was largely due to too small values taken for the octupolar moment
of the transition-metal atom, largely generated by too small
value for $\langle{r_{d}^{4}}\rangle$. Our many-electron
crystal-field based approach enables successful calculations of
the electronic structure and magnetic properties both in the
paramagnetic and in magnetically-ordered state as well as
zero-temperature properties and thermodynamics.

\end{abstract}
\maketitle \vspace {-0.3 cm}

During last 10 years \cite{1,2,3,4} we have described the
magnetism and low-energy electronic structure of NiO within the
localized atomistic paradigm with only three parameters: the
octupolar crystal-field parameter $B_{4}$ =+21 K (10Dq=1.086 eV),
the spin-orbit coupling $\lambda_{s-o}$ =-480 K and a small
trigonal distortion $B_{2}^{0}$ =+50 K \cite{5}. For the magnetic
state the fourth parameter, the molecular-field coefficient $n$,
has been introduced - $n$ = $-200~ T/\mu_{B}$ yields $T_{N}$ of
525 K in agreement with the experimental observation. The
fundamentally important value of $B_{4}$ has been fitted to the
excitation of 1.06-1.13 eV seen in the optical absorption
\cite{6}. Here we would like to calculate $B_{4}$ from first
principles - it buckles the internal consistency of our
understanding of NiO, explaining its insulating ground state,
magnetism both in the AF and paramagnetic state.

Although the ionic crystal-field based approach seems to be the
most natural approach to transition-atom oxides and is known for
75 years starting with pioneering works of Bethe and Van Vleck
\cite {7} the presently in fashion are theoretical approaches
pointing out the itinerant (band) character of $d$ electrons
\cite{8,9,10,11,12,13,14}. This paper has been motivated by a
series of {\it ab initio} and first-principles papers devoted to
explanation of properties of NiO, which recently have been
published within the band picture and the continuous energy
spectrum \cite{15,16,17,18,19,20}. We note that with time the
original LDA approach is significantly modified by incorporation
of different local potentials, i.e. goes towards the advocating
by us many-electron crystal-field approach.

\begin{figure}\vspace {-0.8 cm}
\begin{center}
\includegraphics[width =5.5 cm]{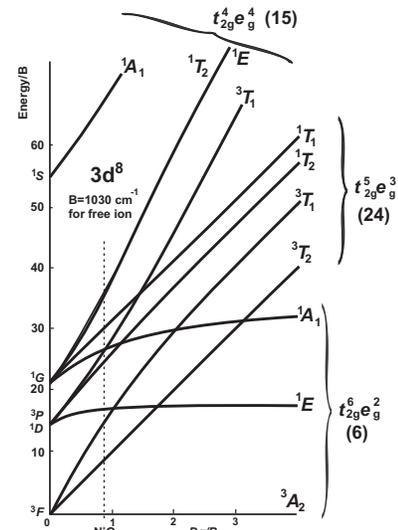}
\end{center} \vspace {-0.8 cm}
\caption{A Tanabe-Sugano diagram for the Ni$^{2+}$ ion (3$d^{8}$
configuration) showing the effect of the octahedral crystal field
on the electronic terms of the free Ni$^{2+}$-ion. The electronic
structure of cubic subterms, corresponding to $Dq$/$B$ =0.85,
relevant to NiO is marked by an arrow. Zero-field term values are
presently well known in the NIST database. \vspace {-0.6 cm}}
\end{figure}

From historical point of view, the band description has been
introduced as an {\bf opposite} view to the ionic view of Van
Vleck, which underlies the discrete electronic structure. After
works of Van Vleck a milestone achievement within the localized
paradigm were the Tanabe-Sugano diagrams known from year of 1954
\cite{21,22} and shown in Fig. 1 for the Ni$^{2+}$ ion. A quite
successful description in 1984 by Fujimori and Minami \cite{23} of
the photoemission data by a local configuration-interaction
cluster model was very soon overflowed by the band-based view
pointing out the dominant role played by the covalency and strong
hybridization of Ni 3$d$ and O $2p$ states.

In the present situation of the dominant band view in the NiO
problem, in general in 3$d$ oxides, we feel necessary to complete
our understanding of properties of the magnetism and electronic
structure of NiO in the localized paradigm. Our understanding is
based on well-known physical concepts like the crystal-field
(CEF) \cite{24}, spin-orbit (s-o) coupling, local distortions and
other terms known from the ionic language. We present this view
being aware that the ionic picture and crystal-field
considerations are at present treated as the "old-fashioned" and
contemptuous physics in times of wide spreading omnipotent band
theories of different versions LDA, LSDA, LDA+U \cite{9,13},
LDA+GGA \cite{12}, LDA+DMFT \cite{20} and many, many others. We
gave a name of QUASST for our approach to a solid containing
transition-metal atoms from Quantum Atomistic Solid State Theory
pointing out that the physically adequate description of
properties of a 3d/4f/5f solid the best is to start from analysis
of the electronic structure of constituting atoms. In a solid
atoms occur rather as ions by which we understand a well-defined
charge state of the given atom.
\begin{figure}[t]\vspace {-0.5cm}
\begin{center}
\includegraphics[width = 5.7 cm]{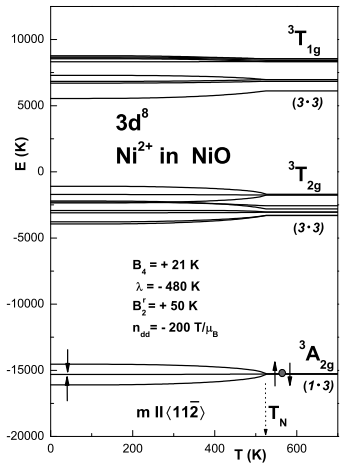}
\end{center} \vspace {-1.1cm}
\caption{The calculated temperature dependence of the $^{3}F$ term
(Ni$^{2+}$, 3$d^{8}$) in NiO in the presence of the octahedral
crystal field and spin-orbit coupling in the magnetically-ordered
state below T$_{N}$ of 525 K. In the paramagnetic state the
electronic structure is temperature independent unless we do
consider, for instance, a changing of the CEF parameter due to
the thermal lattice expansion. The states of the $^{3}A_{2g}$
subterm are characterized by magnetic moment of 0, +2.26 and
-2.26 $\mu_{B}$. In the magnetic state there occurs the spin
polarization of this three-fold degenerated $^{3}A_{2g}$ subterm.
A small splitting, 1.0 meV, of the $^{3}A_{2g}$ subterm in the
paramagnetic state due to the presence of the trigonal distortion
is not visible in this energy scale. } \vspace {-0.5cm}
\end{figure}

Schematic steps of the QUASST approach to NiO can be written as:

1. we accept the atomistic structure of matter, what means that
atoms preserve much of their atomic properties becoming the full
part of a solid. In NiO during the formation of the compound there
occurs an electron transfer of two electrons from Ni to O with
the simultaneous formation of the rocksalt NaCl lattice with the
ionic charge distribution Ni$^{2+}$O$^{2-}$;

2. all magnetism of NiO is related to the Ni$^{2+}$ ion with
eight electrons outside close configuration $^{18}$Ar because
O$^{2-}$ ions have closed shells ($^{10}$Ne);

3. eight electrons of the Ni$^{2+}$ ion form strongly-correlated
atomic-like system, 3d$^{8}$; its electronic structure is known
from the atomic physics (NIST \cite{25}: $^{1}D$ - 1.74 eV,
$^{3}P$ - 2.10, $^{1}G$ - 2.86; $^{1}S$ - 6.51 eV with the
$^{3}F$ Hund rule ground term);

4. from the translational symmetry of the rocksalt structure of
NiO we learn that a) all Ni sites are equivalent, b) the local
symmetry of the Ni cation is octahedral, c) the Ni ion is
surrounded by six nearest neighbor oxygens forming almost perfect
octahedron, d) with the lowering temperature there appears small
off-octahedral trigonal distortion;

5. influence of the octahedral crystal field on the free-ion
electronic terms as is shown in the Tanabe-Sugano diagram, Fig. 1;

6. influence of the intra-atomic spin-orbit coupling on the local
electronic structure; for the free Ni$^{2+}$ ion $\lambda$ = -42
meV; NIST \cite{25} gives the multiplet splitting $^{3}F$ - 0.0
($^{3}F_{4}$), 168.7 meV ($^{3}F_{3}$) and 281.4 meV
($^{3}F_{2}$);

7. influence of the off-octahedral rhombohedral distortion on the
local electronic structure;

8. magnetic interactions lead below 525 K to the
magnetically-ordered state;

9. having determined the local electronic structure with the
eigenfunctions we have the zero-temperature properties as well as
the free energy $F$(T);

10. the zero-temperature properties are related to the properties
of the local ground state; the free energy $F$(T) enables
calculations of the whole thermodynamics. See, for instance, our
description of FeBr$_{2}$ \cite{3}, an exemplary of the magnetic
3$d$ compound.

Ad. 1. An assumption about the atomistic construction of matter
seems to be obvious but we would like to point out that
band-structure calculations for 3$d$ oxides disintegrate 3$d$
atoms completely starting from consideration of all 3$d$ electrons
as independent electrons in the octahedral crystal field.
According to us these one-electron calculations have to
reproduce, before calculations of properties of a solid, the
electronic structure of the given ion, i.e. its term structure.
For NiO, the $^{3}A_{2g}$, $^{1}E$ and $^{1}A_{1}$ subterm
structure of the 6-fold degenerated $t_{2g}^{6}e_{g}^{2}$
configuration. In our approach we simply accept the term
electronic structure known from the atomic physics (see Ad. 3).
We point out that the intraatomic term structure results from
strong electron correlations among $d$ electrons.

Ad. 3. From the atomic physics we know that for the 3d$^{8}$
configuration of the Ni$^{2+}$ ion 45 states are grouped in 5
terms: $^{3}F$ (21 states),  $^{3}P$ (9),  $^{1}G$ (9), $^{1}D$
(5), $^{1}S$ (1). The first excited term is almost 2 eV above
being inactive for low- and room-temperature properties.

Ad. 4. For more complex structures there will be a few 3$d$ sites,
each of them having own electronic structure. Analysis of
low-temperature structure is important to decide if all Ni sites
are equivalent and all ions equally contribute to macroscopic
properties.

Ad. 5. Influence of the octahedral crystal field on the free-ion
electronic terms has been calculated by Tanabe and Sugano in a
year of 1954 already \cite {21,22}. The splitting of electronic
terms to octahedral subterms in a function of the strength of the
octahedral CEF parameter Dq/B (B - intra-ionic Racah parameter
determines the energy scale) is known as Tanabe-Sugano diagrams,
Fig. 1. A problem was and still is with i) the acceptance of its
validity for a solid compound and with ii) the evaluation of the
value of Dq/B on this diagram for a given compound. Numerous
qualitative indications, starting already at fifties of the XX
century, have not been conclusive. In the crystal-field theory
parameter 10Dq ($\cong$600$\cdot$$B_{4}$) in the simplest form is
the multiplication of the octupolar charge moment of the lattice
$A_{4}$ and of the octupolar charge moment of the involved cation
caused by anisotropic charge distribution of the own incomplete
3$d$ shell. Thus Dq ($B_{4}$) can be calculated from first
principles provided the octupolar charge moment, $\beta$ $
\langle{r_{d}^{4}}\rangle$, of the involved ion is known. $\beta$
is the fourth-order Stevens coefficient.

The octahedral crystal field coefficient $A_{4}$ that is the
octupolar charge moment of all surrounding charges at the Ni site
can be calculated from the point-charge model which is
first-principles elementary calculations. Taking the charge of
oxygen Z as -2$\mid$e$\mid$ and the nearest octahedron with the
cation-oxygen distance d of 208.5 pm in NiO we obtain by formulae
$A_{4}$=7$\cdot$Ze$^{2}$/16$\cdot$d$^{5}$ \cite{24} a value for
$A_{4}$ of +290 Ka$_{B}^{-4}$, a$_{B}$ is the Bohr radius. Taking
for the Ni$^{2+}$ ion $\beta$ = +2/315 and a widely-spread
Hartree-Fock result by Freeman-Watson (F-W) $
\langle{r_{d}^{4}}\rangle$ =3.003 a$_{B}^{4}$ \cite {26} a value
of $B_{4}$ of +5.5 K is obtained only. This value is 3.8 times
smaller than the recent experimental determination of $B_{4}$ of
+21 K \cite {5}. In our papers for 3$d$ oxides (LaMnO$_{3}$,
FeBr$_{2}$, CoO, ...) we pointed out that despite this difference
the most important is that these ionic {\it ab initio}
calculations give the proper sign of the $B_{4}$ parameter - it
is important because it determines the ground state of the cation
and the experimentally derived strength of crystal-field
interactions turns out to be much weaker than it was thought so
far in literature for justification of the strong crystal-field
approach in which the one-electron approach becomes more
physically adequate. Here we would like to explain this
theoretical 3.8-times smaller value - we attribute it to a large
underestimation of $\langle{r_{d}^{4}}\rangle$ =3.003 a$_{B}^{4}$
obtained in F-W Hartree-Fock calculations of in a year of 1965
\cite {26}.

At first, we have calculated the contribution to the octupolar
lattice potential from next neighbors - it strongly decreases due
to the increase of the Ni-O distance. The next-nearest
contribution, due to the Ni$^{2+}$ ions, amounts to $9\%$ of the
first-octahedron contribution but it adds making theoretical
evaluation of $A_{4}$ as +315 Ka$_{B}^{-4}$. Thus, we claim that
the proper value of $\langle{r_{d}^{4}}\rangle$ for the Ni$^{2+}$
ion is 10.50 a$_{B}^{4}$ - it is 3.5 times larger value than the
F-W Hartree-Fock result from 1965. We claim that this
widely-spread early Hartree-Fock result is erroneous. Indeed, the
presently derived values are significantly larger. Korotin {\em
et~al.} \cite {27} have used a value for $r_{d}$ of 1.26 $\AA$
yielding approximately $\langle{r_{d}^{4}}\rangle$ = 32.2
a$_{B}^{4}$. Recently Solovyev \cite {28}, page 5, has calculated
for the Ti$^{3+}$ ion $\langle{r_{d}^{2}}\rangle$ of 2.27
$\AA$$^{2}$ (=8.11 a$_{B}^{2}$), which approximately yields
$\langle{r_{d}^{4}}\rangle$ even as large as 65 a$_{B}^{4}$. Thus
we think that a value of 10.5 a$_{B}^{4}$ needed for the purely
ionic electrostatic-origin of the crystal-field splitting in NiO
is fully reasonable.

Ad. 6. It is fundamentally important to take the spin-orbit
coupling into account despite its weakness. Its consequence is
that the physically adequate space becomes the spin-orbital space
not the orbital space separated from the spin space. It
generates, for instance, physically adequate electronic structure
of $d$ electrons and the orbital moment.

Ad. 7. The rhombohedral (trigonal) distortion is small and causes
a slight splitting of the lowest quasi-triplet in NiO with $D$ of
order of a few meV (for $B_{2}^{0}$ value of +50 K shown in Fig.
2 D amounts to 1.0 meV only). But this small distortion determines
in our calculations the direction of the Ni magnetic moment, in
NiO the $\langle{11\bar{2}}\rangle$ direction, i.e.
perpendicularly to the trigonal axis. Such moment direction causes
further symmetry lowering due to magnetostriction. Thus, we can
say that we fully understand, and we calculate, a delicate
interplay of the magnetism of NiO and distortions of its crystal
structure.

Ad. 8. The formation of the magnetic state we have described in
numerous compounds - let mention exemplary 4f/3d/5f compounds
ErNi$_{5}$, FeBr$_{2}$, NiO, CoO, UPd$_{2}$Al$_{3}$ and
UGa$_{2}$, results of which have been published starting from
1992. In all these cases the magnetic energy is much smaller than
the overall CEF splitting. They are both ionic (FeBr$_{2}$
\cite{3}, NiO, CoO \cite{2}) and intermetallic (ErNi$_{5}$,
UPd$_{2}$Al$_{3}$, UGa$_{2}$) \cite{30} compounds.

\begin{figure}[t]\vspace {-0.45cm}
\begin{center}
\includegraphics[width = 11.0 cm]{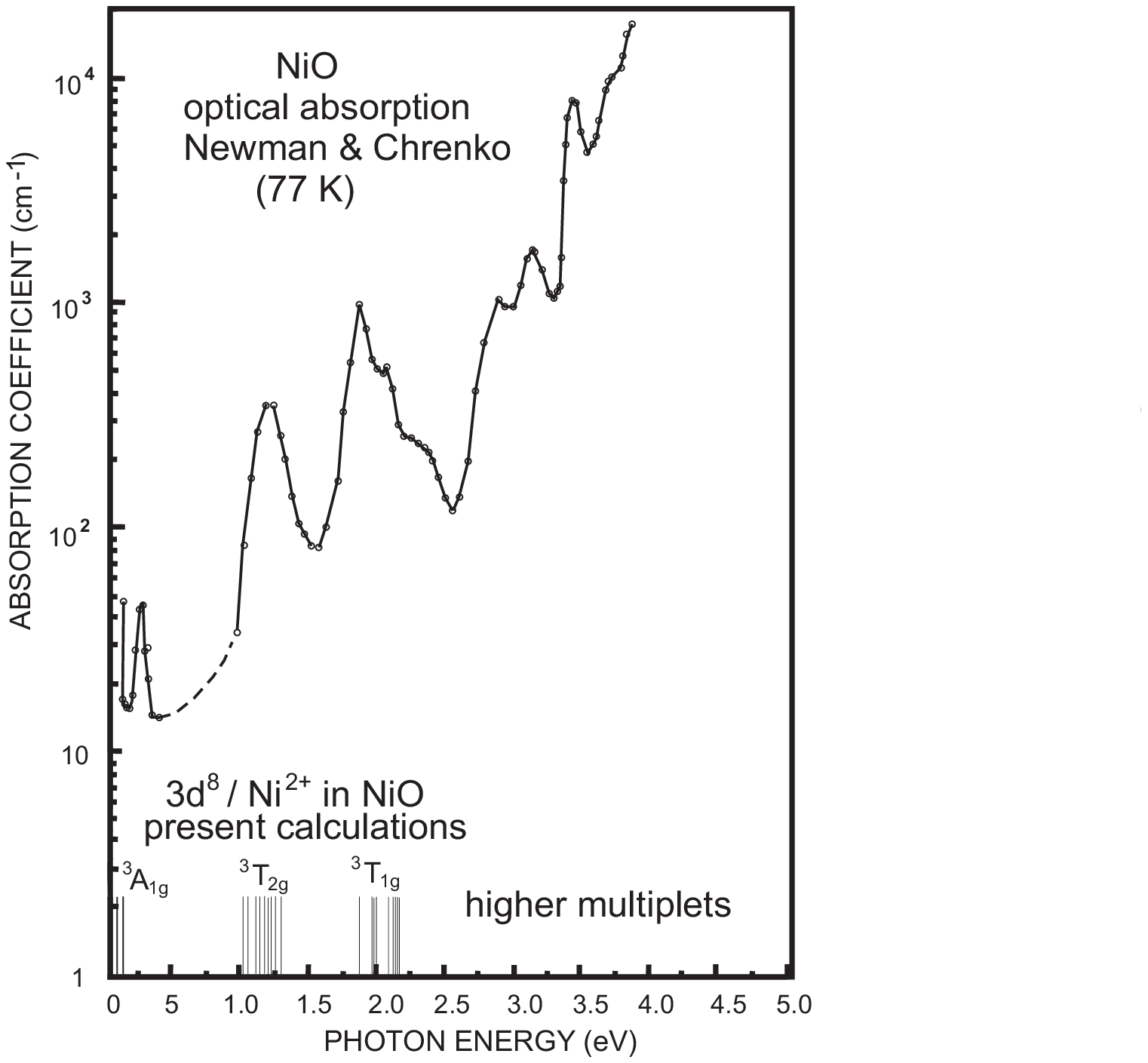}
\end{center}\vspace {-0.7cm}
\caption{Comparison of the calculated electronic structure, shown
in Fig. 2, with that experimentally observed by means of optical
absorption in Ref. \cite{6}. }\vspace {-0.7cm}
\end{figure}
We would like to note that all of the used by us parameters
(dominant octahedral CEF parameter $B_{4}$, the spin-orbit
coupling $\lambda_{s-o}$, lattice distortions) have clear physical
meaning. The most important assumption is the existence of very
strong correlations among 3$d$ electrons preserving the atomistic
ionic integrity of the Ni$^{2+}$ ion also in the solid NiO. We
stress the good reproduction of experimental results like
temperature dependence of the heat capacity \cite{5}, the value
of the magnetic moment and its $\langle{11\bar{2}}\rangle$
direction. Our calculations yield a total magnetic moment of 2.53
$\mu_{B}$ \cite {5} in very good agreement with recent
experimental evaluation \cite {31,32}. The calculated moment
contains almost 20\% contribution from the orbital moment. Our
purely ionic crystal-field approach, despite its simplicity, is
in agreement with recent experimental findings of Jauch and
Reehuis \cite {33}, who have revealed by the high-accuracy single
crystal $\gamma$-ray diffraction experiment the electron-density
distribution in NiO indicating i) the absence of the covalency
and ii) the Ni-O interaction to be purely ionic. For
photoemission results we largely accept the Fujimori-Minami
interpretation \cite{23} as their cluster result about the
multiplet structure, shown in their Fig. 7, is quite similar to
our purely ionic multiplet structure, shown in Figs 1 and 3.

{\bf In conclusions}, we have calculated from first-principles
the octupolar interactions of the Ni$^{2+}$ ion in NiO, which
gives the theoretical basis for the ionic description of
properties of NiO with fully localized strongly-correlated eight
$d$ electrons. The crystal-field interactions are relatively
strong in NiO but not so strong to destroy the ionic integrity of
the 3$d$ electrons. The insulating gap in NiO of 4.3 eV is of the
same origin as a gap in classical insulators, MgO or CaO. A
failure of the up-now first principles ionic calculations for NiO
were largely due to too small values taken for the octupolar
moment of the Ni$^{2+}$ ion, largely generated by too small value
for $\langle{r_{d}^{4}}\rangle$. Our many-electron crystal-field
based approach enables successful calculations of the electronic
structure and magnetic properties both in the paramagnetic and in
magnetically-ordered state as well as zero-temperature properties
and thermodynamics. Thus we claim that this many-electron CEF
picture is the physically adequate starting point for description
of magnetism and low-energy electronic structure of 3$d$ oxides.
Paraphrasing conclusion of Ref. \cite{20} we can say, that in
spite of the limitations of the implementation of magnetism and
spin-polarization we showed that the ionic CEF-based approach
which combines first-principles, material-specific information
with very strong electron correlation effects is able to deal
with late transition-metal monoxides like NiO. This old-known
compound turns out to exhibit subpicosecond switching
characteristics of potential applications in spintronics
\cite{34}. 

\end{document}